\documentstyle[12pt]{article}
\begin{document}
\baselineskip 18pt
\title{Vibrations and Berry Phases of Charged Buckminsterfullerene}
\author{Assa Auerbach \\ Physics
Department, Technion-IIT, Haifa 32000, Israel\thanks{Email:
assa@phassa.technion.ac.il}.}
\maketitle
\begin{abstract}
A simple model of electron-vibron interactions in buckminsterfullerene ions  is
solved semiclassically. Electronic degeneracies of C$_{60}$$^{n-}$ induce
dynamical Jahn-Teller distortions, which are unimodal for $n\!\ne\!3$ and
bimodal for $n\!=\!3$. The quantization of motion along the Jahn-Teller
manifold leads to a symmetric-top rotator Hamiltonian. I find  Molecular
Aharonov-Bohm effects
where electronic Berry phases determine the vibrational spectra, zero point
fluctuations, and electrons' pair binding energies. The latter are relevant to
superconductivity in alkali-fullerenes.
\end{abstract}


For polyatomic molecules, the adiabatic approximation is often used to
eliminate
fast electrons in favor of an effective potential for the
slow nuclei.
This approximation requires special care when the positions of the nuclear
coordinates are near points  of electronic degeneracy. If the electron-ion
interaction is linear  in the ionic displacements (a generic case for
symmetric, non colinear molecules
\cite{ll}) the classical Jahn Teller theory \cite{jt} predicts that the
molecule distorts and some (or  all) of the electronic
degeneracy is lifted. The classical JT theory is controlled by the largeness of
$S\!=\!  |E^{JT}| /(\hbar \omega) $, where $\omega$ is the
characteristic vibrational frequency, and $E^{JT}$ is the relaxed energy of the
distortion. For
$S\!=\!\infty$, (i.e. ``strong coupling'' or ``classical'' limit),  the zero
point motion of the
ions is ignored.

At finite $S$ however,  quantum corrections can be quantitatively and
qualitatively important.  For example, while the JT distortion may break the
Hamiltonian symmetry, quantum fluctuations along the degenerate manifold  or
tunneling between JT minima  can restore the ground state symmetry\cite{lh}.
This is often called the ``Dynamical Jahn Teller'' effect.
In addition, ion coordinates may be subject to quantum interference effects.
Longuet-Higgins has found that a vibrational orbit which surrounds a point
of two-fold electronic degeneracy, can acquire a negative sign from the
transport of
the electron's wave
function \cite{lh}. This effect, often dubbed as
the ``Molecular Aharonov Bohm (MAB) effect'', produces in triangular molecules
half-odd integer quantum numbers. This has been recently confirmed
spectroscopically in Na$_3$ \cite{exp}.   The MAB effect is a simple example of
the geometrical Berry phase, which appears in
a wide host of quantum phenomena \cite{w}.

The soccer-ball shaped molecule C$_{60}$ (buckminsterfullerene)  and its
various crystalline compounds  have ignited enormous interest in the chemistry
and physics community
in past two years \cite{hebard}. Since the discovery of superconductivity
in A$_3$C$_{60}$ (A=K,Cs,Rb), with relatively high transition temperatures
($T_c\approx20^\circ$--$30^\circ K$), much attention has been given to the
electronic properties of  charged  C$_{60}$$^{n-}$ ions.
C$_{60}$ a highly symmetrical molecule (a truncated icosahedron), and its
electronic lowest
unoccupied molecular orbitals (LUMO)  are three-fold degenerate. Thus
the   C$_{60}$$^{n-}$  ions  are natural
candidates for manifestations of dynamical JT effects and MAB effects discussed
above. Several groups have identified the five-fold
degenerate $H_g$ (d-wave like) vibrational modes that couple strongly to the
LUMO orbitals\cite{vzr,sch,t1}. Varma,
Zaanen and Raghavachari  (VZR) \cite{vzr}  proposed that these modes undergo a
dynamical JT
distortion  and calculated the
JT induced pair binding energies at several fillings. These results were used
to explain the large  $T_c$'s of fullerenes relative to doped graphite
superconductors. VZR used the classical approximation, and restricted their
calculation to unimodal distortions (defined later).

Density functional and deformation potential calculations for
C$_{60}$$^{-}$\cite{vzr,sch},  estimate  $E^{JT}\! \approx\!  40{~\rm meV}$.
The important vibrations are in the range of  $\hbar\omega\!\approx $0.1--0.2
eV. Thus, the classical parameter is in fact quite small: $S\!\simeq\!$
0.2--0.4. This indicates that the  ions'
quantum fluctuations  cannot be justifiably neglected.

In this paper the vibrations about dynamically distorted buckminsterfullerine
ions are quantized semiclassically.  I extend previous work of O'Brien, who has
solved the
$n\!=\!1$ case both exactly and semiclassically \cite{ob1,ob2}.
First, the unrestricted classical JT distortions are determined.  For
C$_{60}$$^{n-}$, $n\!\ne \!3$,  the
JT distortions are {\em unimodal}, i.e. involve one quadrupolar mode in the
principle axes frame. For C$_{60}$$^{3-}$, the JT distortion is found to be
{\em bimodal}, i.e. two modes are distorted  simultaneously.
Subsequently, the quantum dynamics parallel and perpendicular  to the JT
manifold are determined.  The excitation spectra and pair binding energies for
$n\!=\!1,\ldots 5$ are
determined up to second order in $S^{-1}$. I will show that that Berry phases
give rise to selection rules for the
pseudo-rotational quantum numbers. These {\em kinematical} restrictions effect
the pairing interaction between electrons, and, therefore, also the
superconducting transition temperature.

This discussion is restricted to the simplest electron-vibron interaction model
of C$_{60}$, which captures the symmetries and degeneracies of this system.
Electron-electron interactions
are presently ignored.   The wave functions of the LUMO $t_{1u}$ states are
represented
by the L=1 triplet $|x\rangle,|y\rangle,|z\rangle$. A single vibronic $H_g$
multiplet is represented by five
real coefficients  \cite{ob1,vzr},
\begin{eqnarray}
q_{m}&=&\sqrt{
\pi/5} \sum_{\mu=-2}^2 M_{m\mu} a_{2\mu}\nonumber\\
M_{m\!\ne\!0,\mu}&=& (2~ {\rm sign} (m) )^{-\frac{1}{2}}
 \left(\delta_{m,\mu} +
{}~{\rm sign} (m)\delta_{m,-\mu}\right),~M_{0,\mu}~=\delta_{\mu,0}
\label{0}
\end{eqnarray}
where $m,\mu\!=\!-2,-1,\ldots 2$, and
$~a_{lm}$ are the coefficients of the spherical harmonics $Y_{lm}$ \cite{ed}.
The Hamiltonian is  $H=H^{el}+H^{vib}$, where
\begin{eqnarray}
H^{el} &=& g {\hbar\omega\over 2}   \pmatrix{ q_{0}-\sqrt{3}
q_{2}&-\sqrt{3}q_{-2}&-\sqrt{3}q_{1} \cr
-\sqrt{3}q_{-2}& q_{0}+\sqrt{3}q_{2}& -\sqrt{3}q_{-1}\cr
-\sqrt{3}q_{1}& -\sqrt{3}q_{-1}& -2 q_{0}},
 \nonumber\\
H^{vib} &=&{ \hbar \omega\over 2} \sum_{\mu }( -   \partial_\mu^2   +   q_{\mu
}^2),
\label{2}
\end{eqnarray}
where $g$ is the dimensionless electron phonon coupling constant.
$H^{vib}$ is invariant under rotations of ${\vec q}$ in $R^5$  and
by construction, the eigenvalues of $H$ are invariant under O(3)
rotations of the molecule's reference frame.
$H^{el}$  is diagonalized by \cite{ob2}
\begin{eqnarray}
H^{el}&=&   ~ g {\hbar\omega\over 2}   T^{-1}(\varpi)~ \pmatrix{z - \sqrt{3}r
&0&0\cr
0& z + \sqrt{3}r   &0\cr
0&0&-2z } ~T(\varpi) \nonumber\\
T&=&   \pmatrix{ \cos \psi & \sin \psi& 0\cr
-\sin \psi& \cos \psi &0\cr
0&0&1}
  \pmatrix{ \cos\theta &0& \sin\theta\cr
0&1&0\cr
\sin\theta&0&\cos\theta}
  \pmatrix{ \cos\phi &  \sin\phi&0\cr
-\sin\phi& \cos\phi & 0\cr
0&0&1}\nonumber\\
\label{4}
\end{eqnarray}
where the Euler angles $\varpi=(\phi,\theta,\psi)$ define the O(3) rotation
which diagonalizes $H^{el}$. In the diagonal basis,
the only non-zero vibrational components which couple to the electrons are
${\vec q}^{~(0)} =(r,0,z,0,0)$.

Since $|{\vec q}|^2$ is invariant under $O(3)$ rotations, the total adiabatic
potential is
\begin{equation}
V(z, r)~= {\hbar\omega\over 2} ( z^2+r^2)~+  {\hbar\omega g\over 2}\left(
n_1(z-\sqrt{3}r)
+n_2(z+\sqrt{3}r) - n_3 2 z \right)  .
\label{4.1}
\end{equation}
$ n_i$ are the occupations of the orbitals $|i\rangle,i=1,2,3$  (ordered from
top to bottom in $H^{el}$), and $\sum_i n_i=n$. $V$ is minimized by the JT
configurations
$({\bar z},{\bar r},{\bar n}_i)$,
which yield the classical energies $E_n^{cl}\!=\!V({\bar z}_n,{\bar r}_n)$. The
distortions are shown in Table I. We choose $S\!=\!\frac{1}{2} g^2$ as our
semiclassical
parameter.

By (\ref{0}), if we define axis $3$ to be at $\tilde\theta\!=\!0$,
the $z$ mode is described by  $  {\bar z}{1\over
2}(3\cos^2{\tilde\theta}\!=\!1)$, and the $r$ mode by
${\bar r}{\sqrt{3}\over 2} \sin^2{\tilde\theta}\cos(2{\tilde\phi})$.
Thus by Table I, $n\!=\!1,2,4,5$ have unimodal distortions  which are symmetric
about the $3$
axis, and   $n\!=\!3$ has  a  bimodal  distortion, about the $3$ and $1$ axes.

In order to quantize the vibrations, it is useful to express  the
kinetic energy
in terms of small fluctuations about the JT distortion. To that end, we
parametrize the  JT degenerate manifold, $\{{\bar q}_\mu\}$, in terms of the
Euler angles $\varpi$ of Eq. (\ref{4}):
\begin{equation}
{\bar q}_\mu(\varpi)~= \sum_{m,m',\mu'=-2}^2 M_{\mu,m} D^{(2)}_{m,m'}(\varpi)
M_{m'\mu'}^{-1} {\bar q}^{(0)}_{\mu'} .
\label{5}
\end{equation}
$D^{(L)}$ is the irreducible rotational matrix of angular momentum $L$
\cite{ed}.
The classical kinetic energy can be derived from (\ref{5}) by the chain rule
for differentiation.  After some cumbersome,
but straightforward, algebra  we obtain the compact expression:
\begin{eqnarray}
\frac{1}{2} |{\dot {\vec q}}|^2&=&
 \frac{1}{2}\left( {\dot z}^2 + {\dot r}^2 +  \sum_{i=1,3} I_i \omega_i^2
\right),\nonumber\\
\omega_1 &=& -\sin \psi {\dot\theta} + \cos\theta\sin\psi
{\dot \phi} ,\nonumber\\
\omega_2 &=& \cos \psi {\dot\theta} + \sin\psi\sin\theta
{\dot \phi},\nonumber\\
\omega_3 &=&{\dot\psi} +\cos\theta{\dot\phi} \nonumber\\
(I_1,I_2,I_3)&=& \left( (\sqrt{3}  z+ r)^2, (\sqrt{3}  z- r)^2 ,4 r^2\right).
\label{6}
\end{eqnarray}
For finite JT distortions, we can identify  $I_i({\bar z},{\bar r})$ as moments
of inertia in the principle axes frame \cite{ed}. Thus, the Euler angles
dynamics follow those of a {\em rigid rotator} \cite{com}.

The unimodal and bimodal cases will be discussed separately. For the unimodal
cases,
${\bar r}\!=\!0$  and $I_i={\bar z}^2 (3,3,0)$ on the JT manifold. The
coordinate $\psi$ decouples from the rotational kinetic, which becomes that of
a
point particle on the sphere. The quantization of the rotational part is
therefore simply $(2I_1)^{-1} {\vec L}^2$ \cite{com}. The remaining coordinates
are  three harmonic oscillators
$r_\gamma\!=\!(r\cos(2\psi ),r\sin(2\psi ),z-{\bar z})$.  Including the
quadratic potential terms in $V(z,r)$, we arrive at the vibrational
eigenenergies:
\begin{equation}
E_n^{uni}[L,n_\gamma]={\hbar \omega }\left({1\over 6{\bar z}_n^2 }L(L+1) +
\sum_{\gamma=1}^3
(n_\gamma +\frac{1}{2} )\right)
\label{7}
\end{equation}
The rotational parts of the eigenfunctions are
\begin{equation}
\Psi_{Lm}^{uni}({\bar q}) =  Y_{Lm}(\theta,\phi) ~ \prod_{is}|n_{is}\rangle'
\label{7.1}\end{equation}
where $ |n_{is}\rangle'$
is an electron Fock state in the  principle axes  basis.
The overlap of this Fock state with a Fock state of the stationary basis
is a determinant that contains $n$ factors of $Y_{1 \nu_{is}}$.
Under reflection, $Y_{L
m}(\pi-\theta,\phi+\pi)\!\to\!(-1)^LY_{Lm}(\theta,\phi)$. Therefore
the electronic wave function yields a Berry phase factor of $(-1)^n$ for
rotations
between inverted points on the sphere  which corespond to closed orbits of
${\bar q}$. Due to the invariance under reflection of ${\bar q}$ (and thus the
left hand side of (\ref{7.1})), a selection rule is obtained:
$(-1)^{L+n}\!=\!1$. Thus, the ground state for $n\!=\!1,5$ has  pseudo-angular
momentum $L\!=\!1$ that contributes to the zero point energy.

The analysis of the bimodal case of $n\!=\!3$ proceeds along similar lines.
{}From Eq.(\ref{6}) and Table I, we see that $(I_1,I_2,I_3)\!=\! 3g^2 (
4,1,1)$. Thus, the kinetic energy includes the rotation of a
rigid body with two equal moments of inertia. The quantization of this system
is the quantum symmetric top Hamiltonian. Fortunately, its solution is a
well-known textbook problem (see e.g. Ref.\cite{ll,ed}).
In addition to the rotator, there are two
harmonic oscillators $r_\gamma\!=\!(z\!-\!{\bar z}, r\!-\!{\bar r})$. The
eigenvalues of the bimodal C$_{60}$$^{3-}$ molecule are thus given by
\begin{equation}
E^{bi}={\hbar \omega }\left({1\over 6g^2  }L(L+1)-
 {1\over 8g^2}k^2~+\sum_{\gamma=1}^2
(n_\gamma +\frac{1}{2} )\right)
\label{8}
\end{equation}
where $L$ and $k$ are quantum numbers of $|{\vec L}|^2$ and $L_{1}$
respectively, and $k\le L$.
The rotational part of the eigenfunctions are
\begin{equation}
\Psi^{bi}_{Lmk}[{\bar q}] =   D_{mk}^{(L)} (\varpi)
\prod_{is}|n_{is}\rangle'
\label{8.1}
\end{equation}
$m$ is the eigenvalue of $L_z$, where $z$ is a stationary axis.
In distinction to the unimodal case, there is no single reflection which
fully classifies the symmetry of the wavefunction.  However, one can obtain
negative signs by transporting the electronic ground state  in certain orbits.
We define the rotation of $\pi$ about principle axis $L_i$  as $C_i$.  The
Berry phases associated with these rotations can be read directly from Eq.
(\ref{4}). For example: for $\psi\!\to\!\psi+\pi$ ($C_3$), the states
$|1\rangle$ and $|2\rangle$ get  multiplied by $(-1)$. Since $D^{(L)}_{m,k}$
transform  as $Y_{Lk}$ under $C_i$,  it is easy to determine their sign
factors. The results  are given below:
\begin{eqnarray}
C_1:|1,0,2\rangle'  \to |1,0,2\rangle' &~~~& C_1:D^{(L)}_{m,k}  \to (-1)^k
D^{(L)}_{m,k}
\nonumber\\
C_2:|1,0,2\rangle'  \to -|1,0,2\rangle' &~~~& C_2:D^{(L)}_{m,k}  \to (-1)^{L+k}
D^{(L)}_{m,-k}
\nonumber\\
C_3:|1,0,2\rangle'  \to -|1,0,2\rangle' &~~~& C_3:D^{(L)}_{m,k}  \to  (-1)^{L}
D^{(L)}_{m,-k} ~.
\label{9}
\end{eqnarray}
Clearly, ${\bar q}$, being coefficients of quadrupole distortions, is invariant
under $C_1,C_2,C_3$. Thus, $C_i$ describe closed orbits in $R^5$. In order to
satisfy (\ref{8.1}) and using the degeneracy of $E^{bi}$ for $k\!\to\!-k$,  we
find that {\em $L$ must be odd and $k$ must be even}. In particular, the ground
state
of (\ref{8}) is given by $L\!=\!1$, and $k\!=\!0$.

A relevant quantity for superconductivity is the ``pair binding''
energy\cite{vzr,ck}
\begin{equation}
U_n= E_{n+1}+E_{n-1} -2E_n    ~,
\label{1}
\end{equation}
where $E_n$ are the total ground state energies. The calculation above finds
that
all odd fillings $n\!=\!1,3,5$
have the same
semiclassical pair binding energy $U_{2n+1} = -2S+1-{1\over 3}S^{-1}$.

In Table~I we summarize the results for the vibrational contributions to the
ground state energies and pair binding energies of  C$_{60}$$^{n-}$. The
semiclassical results contain the leading three terms in the $1/S$ expansion.
However, one may rightfully worry about higher order corrections  since
for C$_{60}$ the experimental estimate is $S\!=~$0.2--0.4. In comparing
O4Brien's  exact results for $n\!=\!1$
\cite{ob2} to the semiclassical expression for $E_1$ given in Table 1, we find
that for $S\ge 0.25 $, the error in the semiclassical approximation is bounded
by $0.2\hbar\omega$. For $S \!<\! 0.25$, the semiclassical error diverges
rapidly. Using $S\!=\!0.25$ for buckminsterfullerene, the semiclassical result
is that $U_3$ is dominated by the rotational energy, which {\em enhances} the
pair binding energy from its
classical value by the significant amount of $\hbar\omega/3$.  We must beware
however, that $S\!=\!0.25$ is pushing the semiclassical expansion quite far.
Since
other important
interactions have not been considered here (e.g.  intermolecular hopping and
electron-electron interactions\cite{ck}), I refrain from inferring quantitative
predictions for $T_c$. The results suggest that further investigations of
(\ref{4}) and its extensions would be worthwhile.

In conclusion, I have shown that the dynamical Jahn teller effect in
C$_{60}$$^{n-}$ involves several interesting features. For $n=1,2,4,5$ the
molecule distorts unimodally, giving rise to a pseudo-angular momentum
spectrum, plus three harmonic oscillators. For $n\!=\!3$, there is a bimodal
distortion, which
generates a spectrum of a symmetric top rotator, plus two harmonic oscillators.
The Berry phases of the electronic wave functions have been calculated. They
determine the allowed
pseudo-angular momenta quantum numbers. It would be interesting if further
spectroscopic investigations of C$_{60}$ ions could resolve the special
structure predicted by the Eqs.(\ref{7},\ref{8}).

\subsubsection*{Acknowledgements} I am
indebted to S. Doniach for his insightful suggestion that the MAB effect might
be important in buckminsterfullerene, and to J.E. Avron for valuable
discussions. This paper was supported in part by grants from the US-Israel
Binational Science Foundation,  the Fund for Promotion of Research at the
Technion, and the US Department of Energy  No. DE-FG02-91ER45441.
\vfill\eject

\vfill\eject
\vskip 3in
\centerline{\underbar{\bf Table I}}
\vskip 0.3in
\begin{tabular}{|c|c|c|c|c|} \hline
$n$ & $({\bar z}_n,{\bar r}_n)$ & $({\bar n}_1,{\bar n}_2,{\bar n}_3)$ & $
E_n/(\hbar\omega)$ &$U_n/(\hbar \omega)$ \\ \hline
0 & $(0 ,0)$ & (0,0,0)&  ${5\over 2}$&   \\
1 & $(g ,0)$ (uni)& (0,0,1)&  $-S+{3\over 2}+{1\over 6}S^{-1}$&   $
-2S+1-{1\over 3}S^{-1}  $  \\
2 & $(2g ,0)$ (uni) & (0,0,2) &  $-4S+{3\over 2}  $& $4S-\frac{1}{2}+{1\over
3}S^{-1} $  \\
3 & $({3\over 2} g ,  {\sqrt{3}\over 2} g )$ (bi) & (1,0,2) & $ -3 S+1+{1\over
6}S^{-1}$& $ -2S+1-{1\over 3}S^{-1}  $       \\
4 & $(-2g ,0)$ (uni)  & (2,2,0)  &   $-4S+ {3\over 2} $&
$4S-\frac{1}{2}+{1\over 3}S^{-1}  $  \\
5 & $(-g ,0)$ (uni) & (1,2,2)&  $-S+{3\over 2}+{1\over 6}S^{-1}$&
$-2S+1-{1\over 3}S^{-1} $  \\
6 & $(0 ,0)$ & (2,2,2)& ${5\over 2}$ &  \\
\hline
\end{tabular}


\begin{thebibliography}{99}
\bibitem{com} One should not confuse the pseudo rotations of Eq. (\ref{6}),
which
describe propagating vibrational waves, with real rotations of the molecule.
Those are
assumed  to be frozen out by crystal fields at the temperatures of interest.
\end{thebibliography}
\end{document}